\documentclass[12pt]{article}
  \usepackage{amsfonts}
  \usepackage{amsmath}
\usepackage{amssymb}
\usepackage{amscd}
\usepackage[dvips]{graphicx}
\begin{document}

~~
\bigskip
\bigskip
\begin{center}
%\section*
{\Large {\bf{{{Twisted acceleration-enlarged Newton-Hooke  space-times and breaking classical symmetry
phenomena}}}}}
\end{center}
\bigskip
\bigskip
\bigskip
\begin{center}
{{\large ${\rm {Marcin\;Daszkiewicz}}$}}
\end{center}
\bigskip
\begin{center}
\bigskip

{ ${\rm{Institute\; of\; Theoretical\; Physics}}$}

{ ${\rm{ University\; of\; Wroclaw\; pl.\; Maxa\; Borna\; 9,\;
50-206\; Wroclaw,\; Poland}}$}

{ ${\rm{ e-mail:\; marcin@ift.uni.wroc.pl}}$}

\end{center}
\bigskip
\bigskip
\bigskip
\bigskip
\bigskip
\bigskip
\bigskip
\bigskip
\bigskip
\begin{abstract}
We find the subgroup of classical acceleration-enlarged Newton-Hooke   Hopf algebra which acts covariantly on the
twisted acceleration-enlarged Newton-Hooke  space-times.  The case of  classical acceleration-enlarged
Galilei quantum group is considered as well.
\end{abstract}
\bigskip
\bigskip
\bigskip
\bigskip
\eject

\section{{{Introduction}}}

Presently,  physicists  expect that there exist aberrations from
relativistic kinematics in high  energy  (transplanckian) regime.
Such a suggestion follows from many theoretical
\cite{theo1}, \cite{theo2} as well as experimental
(see e.g. \cite{exp1}) %, \cite{exp2}
investigations performed in the last time.

Generally, there exist two approaches to describe the particle
kinematics in ultra-high energy regime. First of them assumes that
relativistic symmetry becomes broken at Planck's scale to the proper
subgroup of Poincare algebra \cite{vsr1}, \cite{vsr2}. The second
approach is more sofisticated, i.e. it assumes that relativistic
symmetry is still present in high energy regime, but it becomes
deformed \cite{def1}.%, \cite{def2}.

The first treatment has been proposed in \cite{vsr1}, \cite{vsr2}
where authors assumed that the whole Lorentz algebra is broken to
the four subgroups: T(2), E(2),  HOM(2)  and SIM(2)
 identified with
four versions of so-called Very Special Relativity. The second
treatment arises from Quantum Group Theory \cite{qg1}, \cite{qg2}
which, in accordance with Hopf-algebraic classification of all
relativistic and nonrelativistic deformations \cite{class1},
\cite{class2}, provides three types of quantum spaces. First of them
corresponds to the well-known canonical type of noncommutativity
\begin{equation}
[\;{\hat x}_{\mu},{\hat x}_{\nu}\;] =
i\theta_{\mu\nu}\;,%\;\;\;;\;\;\; \theta_{\mu\nu} = {\rm const}\;,
\label{wielkaslawia}
\end{equation}
with antisymmetric constant tensor $\theta^{\mu\nu}$. Its
relativistic and nonrelativistic Hopf-algebraic counterparts have
been proposed in  \cite{chi} and \cite{daszkiewicz} respectively.\\
The second kind of mentioned deformations  introduces the
Lie-algebraic type of space-time noncommutativity
\begin{equation}
[\;{\hat x}_{\mu},{\hat x}_{\nu}\;] = i\theta_{\mu\nu}^{\rho}{\hat
x}_{\rho}\;, \label{noncomm1}
\end{equation}
with particularly chosen coefficients $\theta_{\mu\nu}^{\rho}$ being
constants. The corresponding Poincare quantum groups have been
introduced in \cite{kappaP}-\cite{lie2}, while the suitable Galilei
algebras  - in \cite{kappaG} and \cite{daszkiewicz}.\\
The last kind of quantum space, so-called quadratic type of
noncommutativity
\begin{equation}
[\;{\hat x}_{\mu},{\hat x}_{\nu}\;] =
i\theta_{\mu\nu}^{\rho\tau}{\hat x}_{\rho}{\hat
x}_{\tau}\;\;\;;\;\;\;\theta_{\mu\nu}^{\rho\tau} = {\rm const.}\;,
\label{noncomm2}
\end{equation}
has been  proposed  in \cite{qdef}, \cite{paolo} and \cite{lie2} at
relativistic and in \cite{daszkiewicz2} at nonrelativistic level.

The links between both (mentioned above) approaches have been
investigated recently in articles \cite{link1} and \cite{link2}.
Preciously, it has been demonstrated that the very particular
realizations of canonical, Lie-algebraic and quadratic space-time
noncommutativity are covariant with respect the action of undeformed
T(2), E(2) and HOM(2) %SIM(2)
 subgroups respectively. Such a result
seems to be quite interesting because it connects two different
approaches to the same problem - to the form  of Poincare algebra at
Planck's scale. It also confirms expectation that relativistic
 symmetry in high energy regime should be modified,
while the realizations of such an idea by breaking or deforming of
Poincare  algebra plays only the secondary role.

In this article we extend described above investigations to the case of classical
accelera-\\tion-enlarged Newton-Hooke Hopf algebras $\;{\mathcal U}_{0}(\widehat{NH}_{
\pm})$  \cite{Lukierski:2007ed}, \cite{Gomis:2008jc}. Particulary, we find their subgroups which act covariantly on the
following
(provided in \cite{Daszkiewicz:2010bp} and \cite{Daszkiewicz:2009px}) twist-deformed acceleration-enlarged Newton-Hooke
space-times\footnote{$x_0 = ct$.},\footnote{It should be noted that symbol $\tau$ plays the role of time scale parameter (cosmological constant), which is responsible
for oscillation or expansion of space-time noncommutativity (\ref{ssnhspace}).
For  $\tau$ approaching infinity we reproduce the canonical (\ref{wielkaslawia}),
Lie-algebraic (\ref{noncomm1}) and quadratic (\ref{noncomm2}) type of space-time noncommutativity.}
\begin{equation}
[\;t,{ x}_{i}\;] = 0\;\;\;,\;\;\; [\;{ x}_{i},{ x}_{j}\;] =
if_{\pm}\left(\frac{t}{\tau}\right)
\;, \label{ssnhspace}
\end{equation}
with
$$f_+\left(\frac{t}{\tau}\right) =
f\left(\sinh\left(\frac{t}{\tau}\right),\cosh\left(\frac{t}{\tau}\right)\right)\;\;\;,\;\;\;
f_-\left(\frac{t}{\tau}\right) =
f\left(\sin\left(\frac{t}{\tau}\right),\cos\left(\frac{t}{\tau}\right)\right)\;.$$
Further, by contraction limit of obtained results
$(\tau \to \infty)$, we analyze the case of so-called classical acceleration-enlarged Galilei Hopf
algebra $\;{\mathcal U}_{0}(\widehat{G})$ proposed in \cite{Lukierski:2007nh}.

The paper is organized as follows. In second section we describe the
general algorithm used in present article. Sections three and four are
  devoted respectively  to the  subgroups of classical acceleration-enlarged Newton-Hooke as well as classical
  acceleration-enlarged  Galilei Hopf  symmetries
    acting covariantly on the proper (acceleration-enlarged)  twist-deformed space-times (\ref{ssnhspace}). Final remarks are presented in the last section.

\section{{{General prescription}}}

In this section we describe the general algorithm which can be applied to the arbitrary twist deformation
of  space-time symmetries algebra ${\cal A}$.

First of all, we recall basic facts related with  the twist-deformed
quantum group $\;{\mathcal U}_{\mathcal{F}}({\cal A})$ and with the corresponding quantum space-time.  In
accordance with  general  twist procedure \cite{drin}, the
algebraic sector of   Hopf structure $\;{\mathcal U}_{\mathcal{F}}({\cal A})$ remains undeformed, while
the   coproducts and antipodes  transform as follows
\begin{equation}
\Delta _{0}(a) \to \Delta _{\mathcal{F} }(a) = \mathcal{F}\circ
\,\Delta _{0}(a)\,\circ \mathcal{F}^{-1}\;\;\;,\;\;\;
S_{\mathcal{F}}(a) =u_{\mathcal{F} }\,S_{0}(a)\,u^{-1}_{\mathcal{F}
}\;,\label{fs}
\end{equation}
with $\Delta _{0}(a) = a \otimes 1 + 1 \otimes a$, $S_0(a) = -a$
and $u_{\mathcal{F} }=\sum f_{(1)}S_0(f_{(2)})$ (we use Sweedler's
notation $\mathcal{F}=\sum f_{(1)}\otimes f_{(2)}$). Present in the
above formula twist element $\mathcal{F} \in {\mathcal
U}_{\mathcal{F}}({\cal A}) \otimes {\mathcal U}_{\mathcal{F}}({
\cal A})$ satisfies the classical cocycle condition
\begin{equation}
{\mathcal F}_{12} \cdot(\Delta_{0} \otimes 1) ~{\cal F} = {\mathcal
F}_{23} \cdot(1\otimes \Delta_{0}) ~{\mathcal F}\;, \label{cocyclef}
\end{equation}
and the normalization condition
\begin{equation}
(\epsilon \otimes 1)~{\cal F} = (1 \otimes \epsilon)~{\cal F} = 1\;,
\label{normalizationhh}
\end{equation}
with ${\cal F}_{12} = {\cal F}\otimes 1$, ${\cal F}_{23} = 1 \otimes
{\cal F}$  and
\begin{equation}
\Delta _{0}(a) = a \otimes 1 + 1 \otimes a\;. \label{clasical}
\end{equation}

The corresponding to the above Hopf structure
 space-time is defined as  quantum representation space (Hopf
module)  with action of
the  symmetry generators satisfying suitably deformed  Leibnitz
rules \cite{bloch}, \cite{chi}
\begin{equation}
h\rhd \omega _{\mathcal{F}}\left( f(x)\otimes g(x)\right) =\omega _{\mathcal{%
F}}\left( \Delta _{\mathcal{F}}(h)\rhd f(x)\otimes g(x)\right)\;,
\label{laibnitz}
\end{equation}
for  $h\in \,{\mathcal U}_{\mathcal{F}}({\cal A})$ or $\,{\mathcal
U}_{\mathcal{F}}({\cal A})$  and
\begin{equation}
\omega _{\mathcal{F}}\left( f(x)\otimes g(x)\right)
=\omega\circ\left(
 \mathcal{F}^{-1}\rhd  f(x)\otimes g(x)\right)\;\;\;;\;\;\;\omega\circ\left(
a\otimes b\right) = a\cdot b \;. \label{definition}
\end{equation}
The action of $\;{\mathcal U}_{\mathcal{F}}({\cal A})$ algebra on its Hopf module of
functions depending on space-time coordinates ${x}_\mu$ is given by
\begin{equation}
h\rhd f(x)=h\left(x_\mu,\partial_\mu\right)f(x)\;,  \label{aa1}
\end{equation}
 while the $\star_{\mathcal{F}}$-multiplication of
arbitrary two functions  is defined as follows
\begin{equation}
f({x})\star_{\mathcal{F}} g({x}):= \omega\circ\left(
 \mathcal{F}^{-1}\rhd  f({x})\otimes g({x})\right) \;.
\label{star}
\end{equation}
It should be also noted that the commutation relations
\begin{equation}
[\;{  x}_{\mu},{  x}_{\nu}\;]_{\star_{\mathcal{F}}} = { x}_{\mu}\,
{\star_{\mathcal{F}}}\, {  x}_{\nu} - {  x}_{\nu}\,
{\star_{\mathcal{F}}} \,{  x}_{\mu}
 \;,
\label{slavic}
\end{equation}
are covariant (by definition) with respect the action of Hopf algebra  generators
(see deformed Leibnitz rules
(\ref{laibnitz})).

In this article we consider the action of undeformed acceleration-enlarged
Newton-Hooke  as well as classical acceleration-enlarged Galilei  Hopf algebras on the commutation relations (\ref{slavic}) $({\cal  A} = \widehat{NH}_{\pm}\;{\rm or}\; \widehat{G})$. It
is given by the particular realizations of differential representation (\ref{a1}) and
new classical Leibnitz rules
\begin{equation}
h\rhd \omega_{\mathcal{F}} \left( f(x)\otimes g(x)\right)
=\omega_{\mathcal{F}} \left( \Delta _{0}(h)\rhd f(x)\otimes
g(x)\right)\;, \label{classlaibnitz}
\end{equation}
associated with coproduct (\ref{clasical}). Further, we demonstrate
that in such a case the relations (\ref{slavic}) are not invariant
with respect the action of the whole algebras $\;{\mathcal U}_{0}(\widehat{NH}_{
\pm})$ and $\;{\mathcal U}_{0}(\widehat{ G})$, but only with respect their
proper subgroups. Such an effect can be identified with  the breaking classical symmetry phenomena associated
with twist-deformed space-times (\ref{slavic}).

\section{{{Breaking of classical acceleration-enlarged Newton-Hooke symmetry}}}

In this section we turn to the case of undeformed acceleration-enlarged Newton-Hooke  Hopf
algebra $\;{\mathcal U}_0(\widehat{ NH}_{\pm})$ defined by the following algebraic sector\footnote{The both Hopf structures $\;{\mathcal
U}_0(\widehat{ NH}_{\pm})$ contain, apart from rotation $(M_{ij})$,
boost $(K_{i})$ and space-time translation $(P_{i}, H)$ generators,
the additional ones denoted by $F_{i}$, responsible for constant
acceleration.}
\begin{eqnarray}
&&\left[\, M_{ij},M_{kl}\,\right] =i\left( \delta
_{il}\,M_{jk}-\delta _{jl}\,M_{ik}+\delta _{jk}M_{il}-\delta
_{ik}M_{jl}\right)\;\; \;, \;\;\; \left[\, H,P_i\,\right] =\pm
\frac{i}{\tau^2}K_i
 \;,  \notag \\
&~~&  \cr &&\left[\, M_{ij},K_{k}\,\right] =i\left( \delta
_{jk}\,K_i-\delta _{ik}\,K_j\right)\;\; \;, \;\;\;\left[
\,M_{ij},P_{k }\,\right] =i\left( \delta _{j k }\,P_{i }-\delta _{ik
}\,P_{j }\right) \;,\nonumber
\\
&~~&  \cr &&\left[ \,M_{ij},H\,\right] =\left[ \,K_i,K_j\,\right] =
\left[ \,K_i,P_{j }\,\right] =0\;\;\;,\;\;\;\left[ \,K_i,H\,\right]
=-iP_i\;\;\;,\;\;\;\left[ \,P_{i },P_{j }\,\right] = 0\;,\label{nnnga}\\
&~~&  \cr &&\left[\, F_i,F_j\,\right] =\left[\, F_i,P_j\,\right]
=\left[\, F_i,K_j\,\right] =0\;\; \;, \;\;\;\left[\,
M_{ij},F_{k}\,\right] =i\left( \delta _{jk}\,F_i-\delta
_{ik}\,F_j\right) \;,\nonumber\\
&~~&  \cr &&~~~~~~~~~~~~~~~~~~~~~~~~~~~~~~~~~~~~\left[\,
H,F_{i}\,\right] =2iK_i\;,\nonumber
\end{eqnarray}
and classical coproduct (\ref{clasical}). One can  check that the above structure is  represented on Hopf
module of functions as follows (see formula (\ref{aa1}))
\begin{equation}
H\rhd f(t,\overline{x})=i{\partial_t}f(t,\overline{x})\;\;\;,\;\;\;
P_{i}\rhd f(t,\overline{x})=iC_{\pm} \left(\frac{t}{\tau}\right)
{\partial_i}f(t,\overline{x})\;, \label{a1}
\end{equation}
\begin{equation}
M_{ij}\rhd f(t,\overline{x}) =i\left( x_{i }{\partial_j} -x_{j
}{\partial_i} \right) f(t,\overline{x})\;\;\;,\;\;\; K_i\rhd
f(t,\overline{x}) =i\tau \,S_{\pm} \left(\frac{t}{\tau}\right)
{\partial_i} \,f(t,\overline{x})\;,\label{dsfa}
\end{equation}
and
\begin{equation}
F_i\rhd f(t,\overline{x})=\pm 2i\tau^2\left(C_{\pm}
\left(\frac{t}{\tau}\right) -1\right)
{\partial_i}f(t,\overline{x})\;, \label{dsf}
\end{equation}
with
$$C_{+/-} \left(\frac{t}{\tau}\right) = \cosh/\cos \left(\frac{t}{\tau}\right)\;\;\;{\rm and}\;\;\;
S_{+/-} \left(\frac{t}{\tau}\right) = \sinh/\sin
\left(\frac{t}{\tau}\right) \;.$$

As it was already mentioned in Introduction the twist deformations of  quantum group $\;{\mathcal U}_0(\widehat{ NH}_{\pm})$
have been provided in \cite{Daszkiewicz:2010bp}. Here, we take under consideration
the twisted acceleration-enlarged Newton-Hooke space-times defined by the following twist factors
\begin{eqnarray}
{\cal F} &=& {\cal F}_{\alpha_1} =
\exp \left[\;\frac{i}{4}\sum_{k,l=1}^{2}{\alpha_1^{kl}} P_k \wedge P_l\;\right]\;\;\;\;\;\;\;\,
[\;\alpha_1^{kl} = -\alpha_1^{lk}=\alpha_1\;]\;, \label{macierze01}\\
&~~&\cr {\cal F} &=& {\cal F}_{\alpha_2} = \exp \left[\;\frac{i}{4}\sum_{k,l=1}^{2}\alpha_2^{kl} K_k \wedge
P_l\;\right]\;\;\;\;\;\;\; [\;\alpha_2^{kl} = -\alpha_2^{lk}=\alpha_2\;]\;,\\ &~~&\cr
{\cal F} &=& {\cal F}_{\alpha_3} = \exp \left[\;\frac{i}{4}\sum_{k,l=1}^{2}{\alpha_3^{kl}} K_k \wedge
K_l\;\right]\;\;\;\;\;\;\, [\;\alpha_3^{kl} = -\alpha_3^{lk}=\alpha_3\;]\;,\label{macierzennh}\\
{\cal F} &=&{\cal F}_{\alpha_4} =  \exp \left[\;\frac{i}{4}\sum_{k,l=1}^{2}{\alpha_4^{kl}} F_k \wedge
F_l \;\right]\;\;\;\;\;\;\;\, [\;\alpha_4^{kl} = -\alpha_4^{lk} = \alpha_4\;]\;,
\label{macierze0100}\\&~~&\cr
{\cal F} &=& {\cal F}_{\alpha_5} =
\exp \left[\;\frac{i}{4}\sum_{k,l=1}^{2}\alpha_5^{kl} F_k \wedge P_l\;\right]\;\;\;\;\;\;\; [\;\alpha_5^{kl}
= -\alpha_5^{lk}=\alpha_5\;]\;,\label{macierze100}\\&~~&\cr
{\cal F} &=& {\cal F}_{\alpha_6} = \exp \left[\;\frac{i}{4}\sum_{k,l=1}^{2}{\alpha_6^{kl}} K_k \wedge
F_l\;\right]\;\;\;\;\;\;\, [\;\alpha_6^{kl} = -\alpha_6^{lk}=\alpha_6\;]\;.
\label{macierzenn}
\end{eqnarray}
In other words, we consider  spaces of the form
\begin{equation}
[\;t,{x}_{i}\;]_{\star_{\mathcal{F}}} =[\;{x}_{1},{x}_{3}\;]_{\star_{\mathcal{F}}} = [\;{x}_{2},{x}_{3}\;]_{\star_{\mathcal{F}}} =
0\;\;\;,\;\;\; [\;{x}_{1},{x}_{2}\;]_{\star_{\mathcal{F}}} =
if({t})\;\;;\;\;i=1,2,3
\;, \label{spaces}
\end{equation}
with  function $f({t})$ given by\\
\\
\begin{eqnarray}
f({t})&=&f_{\kappa_1}({t}) =
f_{\pm,\kappa_1}\left(\frac{t}{\tau}\right) = \kappa_1\,C_{\pm}^2
\left(\frac{t}{\tau}\right)\;, \label{w2}\\
f({t})&=&f_{\kappa_2}({t}) =
f_{\pm,\kappa_2}\left(\frac{t}{\tau}\right) =\kappa_2\tau\, C_{\pm}
\left(\frac{t}{\tau}\right)S_{\pm} \left(\frac{t}{\tau}\right) \;,
\label{w3}\\
f({t})&=&f_{\kappa_3}({t}) =
f_{\pm,\kappa_3}\left(\frac{t}{\tau}\right) =\kappa_3\tau^2\,
S_{\pm}^2 \left(\frac{t}{\tau}\right) \;, \label{w4}\\
f({t})&=&f_{\kappa_4}({t}) =
 f_{\pm,\kappa_4}\left(\frac{t}{\tau}\right) = 4\kappa_4
 \tau^4\left(C_{\pm}\left(\frac{t}{\tau}\right)
-1\right)^2 \;, \label{w5}\\
f({t})&=&f_{\kappa_5}({t}) =
f_{\pm,\kappa_5}\left(\frac{t}{\tau}\right) = \pm \kappa_5\tau^2
\left(C_{\pm}\left(\frac{t}{\tau}\right)
-1\right)C_{\pm} \left(\frac{t}{\tau}\right)\;, \label{w6}\\
f({t})&=&f_{\kappa_6}({t}) =
f_{\pm,\kappa_6}\left(\frac{t}{\tau}\right) = \pm \kappa_6\tau^3
\left(C_{\pm}\left(\frac{t}{\tau}\right) -1\right)S_{\pm}
\left(\frac{t}{\tau}\right)\;.\label{w7}
\end{eqnarray}\\
\\
\\
Of course, for all  parameters $\kappa_a$ running to zero the above space-times become commutative.

Let us now turn to the covariance of relations (\ref{w2})-(\ref{w7}) with respect the action of undeformed
Hopf algebra $\;{\mathcal U}_0(\widehat{ NH}_{\pm})$. Using differential representation (\ref{a1})-(\ref{dsf}), classical  Leibnitz rules
(\ref{clasical}) and twist factors (\ref{macierze01})-(\ref{macierzenn}), one finds (see prescription (\ref{classlaibnitz}))
\\
\\
\begin{eqnarray}
&&~~~~~~~~~~~~~~~~~~~G_k \;\rhd\; \left[\;t,{x}_{i}\;\right]_{\star_{\mathcal{F}}}  = 0\;, \label{row1}\\
G_k &\rhd& \left[\;\left[\;x_i,{x}_{j}\;\right]_{\star_{\mathcal{F}}}-if(t)(\delta_{1i}\delta_{2j}- \delta_{1j}\delta_{2i}) \;\right] = 0\;\;;\;\;G_k = P_k,\;K_k,\;F_k\;, \label{row2}\\
M_{kl} &\rhd& \left[\;t,{x}_{i}\;\right]_{\star_{\mathcal{F}}}  = 0\;\;\;,\;\;\;M_{12} \;\rhd \; \left[\;\left[\;x_i,{x}_{j}\;\right]_{\star_{\mathcal{F}}}-if(t)(\delta_{1i}\delta_{2j} - \delta_{1j}\delta_{2i})\;\right] = 0\;, \label{row4}\\
M_{13} &\rhd& \left[\;\left[\;x_i,{x}_{j}\;\right]_{\star_{\mathcal{F}}}-if(t)(\delta_{1i}\delta_{2j} - \delta_{1j}\delta_{2i}) \;\right] = f(t)(\delta_{2i}\delta_{3j}-\delta_{2j}\delta_{3i})\;, \label{row5}\\
M_{23} &\rhd& \left[\;\left[\;x_i,{x}_{j}\;\right]_{\star_{\mathcal{F}}}-if(t)(\delta_{1i}\delta_{2j} - \delta_{1j}\delta_{2i}) \;\right] = -f(t)(\delta_{1i}\delta_{3j}-\delta_{1j}\delta_{3i})\;, \label{row6}\\
H &\rhd& \left[\;\left[\;x_i,{x}_{j}\;\right]_{\star_{\mathcal{F}}}-if(t)(\delta_{1i}\delta_{2j} - \delta_{1j}\delta_{2i}) \;\right] = h(t)(\delta_{1i}\delta_{2j} - \delta_{1j}\delta_{2i})\;,\\
&&~~~~~~~~~~~~~~~~~~~~~~~~~~~~~~~~~~~~~~~~~H  \;\rhd\; \left[\;t,{x}_{i}\;\right]_{\star_{\mathcal{F}}}  = 0\;,\label{row5}
\end{eqnarray}
with $h(t) = \frac{d f(t)}{dt}$, i.e.
\begin{eqnarray}
h({t})&=&h_{\kappa_1}({t}) =
h_{\pm,\kappa_1}\left(\frac{t}{\tau}\right) = \pm\frac{\kappa_1}{\tau}\,S_{\pm}
\left(\frac{2t}{\tau}\right)\;, \label{pw2}\\
h({t})&=&h_{\kappa_2}({t}) =
h_{\pm,\kappa_2}\left(\frac{t}{\tau}\right) =\kappa_2\, C_{\pm}
\left(\frac{2t}{\tau}\right) \;,
\label{pw3}\\
h({t})&=&h_{\kappa_3}({t}) =
h_{\pm,\kappa_3}\left(\frac{t}{\tau}\right) =\kappa_3\tau\,
S_{\pm} \left(\frac{2t}{\tau}\right) \;, \label{pw4}\\
h({t})&=&h_{\kappa_4}({t}) =
 h_{\pm,\kappa_4}\left(\frac{t}{\tau}\right) = \pm 8\kappa_4
 \tau^3S_{\pm}
\left(\frac{t}{\tau}\right)\left(C_{\pm}\left(\frac{t}{\tau}\right)
-1\right) \;, \label{pw5}\\
h({t})&=&h_{\kappa_5}({t}) =
h_{\pm,\kappa_5}\left(\frac{t}{\tau}\right) = \kappa_5\tau
\left(S_{\pm}\left(\frac{2t}{\tau}\right)
-S_{\pm} \left(\frac{t}{\tau}\right)\right)\;, \label{pw6}\\
h({t})&=&h_{\kappa_6}({t}) =
h_{\pm,\kappa_6}\left(\frac{t}{\tau}\right) = 2\kappa_6\tau^2
\left(2C_{\pm}\left(\frac{t}{\tau}\right) +1\right)S_{\pm}^2
\left(\frac{t}{2\tau}\right)\;.\label{pw7}
\end{eqnarray}
The above result means that the commutation relations (\ref{w2})-(\ref{w7}) remain invariant with respect the action
of $P_i$, $K_i$, $F_i$ and $M_{12}$ generators. Hence, the "isometry" condition for considered (twisted) spaces
breaks the whole $\;{\mathcal U}_0(\widehat{ NH}_{\pm})$ quantum group into its subalgebra generated by spatial translations, boosts,
constant acceleration generators and rotation in $(x_1,x_2)$-plane.

Finally, it should be noted that one can easily extend the above algorithm to the case of usual Newton-Hooke Hopf structure
$\;{\mathcal U}_0({ NH}_{\pm})$
by putting acceleration generators $F_i$ equal zero.

\section{{{The case of  acceleration-enlarged Galilei Hopf algebra analyzed  in the contraction limit
$(\tau \to \infty)$ of $\;{\mathcal U}_0(\widehat{ NH}_{\pm})$ Hopf structure}}}

Let us now turn to the classical acceleration-enlarged Galilei Hopf algebra $\;{\mathcal U}_0(\widehat{ G})$  given by the following
algebraic sector
\begin{eqnarray}
&&\left[\, M_{ij},M_{kl}\,\right] =i\left( \delta
_{il}\,M_{jk}-\delta _{jl}\,M_{ik}+\delta _{jk}M_{il}-\delta
_{ik}M_{jl}\right)\;\; \;, \;\;\; \left[\, H,P_i\,\right] =0
 \;,  \notag \\
&~~&  \cr &&\left[\, M_{ij},K_{k}\,\right] =i\left( \delta
_{jk}\,K_i-\delta _{ik}\,K_j\right)\;\; \;, \;\;\;\left[
\,M_{ij},P_{k }\,\right] =i\left( \delta _{j k }\,P_{i }-\delta _{ik
}\,P_{j }\right) \;, \label{gggali}
\\
&~~&  \cr &&\left[ \,M_{ij},H\,\right] =\left[ \,K_i,K_j\,\right] =
\left[ \,K_i,P_{j }\,\right] =0\;\;\;,\;\;\;\left[ \,K_i,H\,\right]
=-iP_i\;\;\;,\;\;\;\left[ \,P_{i },P_{j }\,\right] = 0\;,\nonumber\\
&~~&  \cr &&\left[\, F_i,F_j\,\right] =\left[\, F_i,P_j\,\right]
=\left[\, F_i,K_j\,\right] =0\;\; \;, \;\;\;\left[\,
M_{ij},F_{k}\,\right] =i\left( \delta _{jk}\,F_i-\delta
_{ik}\,F_j\right) \;,\nonumber\\
&~~&  \cr &&~~~~~~~~~~~~~~~~~~~~~~~~~~~~~~~~~~~~\left[\,
H,F_{i}\,\right] =2iK_i\;,\nonumber
\end{eqnarray}
and trivial coproduct (\ref{clasical}). It is well-known that the above Hopf
structure  can be get by the contraction limit $(\tau \to \infty)$ of discussed
in pervious section  quantum group $\;{\mathcal U}_0(\widehat{ NH}_{\pm})$. \\
The noncommutative  space-times associated with twist deformations of  Hopf algebra $\;{\mathcal U}_0(\widehat{G})$ can be
 provided
by the contraction procedure of spaces (\ref{w2})-(\ref{w7}); they take the form
\begin{equation}
[\;t,{x}_{i}\;]_{\star_{\mathcal{F}}} =[\;{x}_{1},{x}_{3}\;]_{\star_{\mathcal{F}}} = [\;{x}_{2},{x}_{3}\;]_{\star_{\mathcal{F}}} =
0\;\;\;,\;\;\; [\;{x}_{1},{x}_{2}\;]_{\star_{\mathcal{F}}} =
iw({t})\;\;;\;\;i=1,2,3
\;, \label{wwspaces}
\end{equation}
with $(w_{\kappa_i}(t) = \lim_{\tau \to \infty} f_{\kappa_i}(t))$
\begin{eqnarray}
w(t) &=& w_{\kappa_1}({t}) = \kappa_1\;,\label{ggnw2}\\
w(t) &=& w_{\kappa_2}({t}) = \kappa_2\,t\;,\label{ggnw3}\\
w(t) &=& w_{\kappa_3}({t}) = \kappa_3\,t^2\;,\label{ggnw4}\\
w(t) &=& w_{\kappa_4}({t}) = \kappa_4\,t^4\;,  \label{ggnw5}\\
w(t) &=& w_{\kappa_5}({t}) = \frac{1}{2}\kappa_5\,t^2\;, \label{ggnw6}\\
w(t) &=& w_{\kappa_6}({t}) = \frac{1}{2}\kappa_6\,t^3\;. \label{ggnw7}
\end{eqnarray}
It should be also noted, that the Galileian counterpart of covariance conditions (\ref{row1})-(\ref{row5}) in $\tau \to \infty$ limit looks
as follows
\begin{eqnarray}
&&~~~~~~~~~~~~~~~~~~~G_k \;\rhd\; \left[\;t,{x}_{i}\;\right]_{\star_{\mathcal{F}}}  = 0\;, \label{ffrow1}\\
G_k &\rhd& \left[\;\left[\;x_i,{x}_{j}\;\right]_{\star_{\mathcal{F}}}-iw(t)(\delta_{1i}\delta_{2j} - \delta_{1j}\delta_{2i})\;\right] = 0\;\;;\;\;G_k = P_k,\;K_k,\;F_k\;, \label{ffrow2}\\
M_{kl} &\rhd& \left[\;t,{x}_{i}\;\right]_{\star_{\mathcal{F}}}  = 0\;\;\;,\;\;\;M_{12} \;\rhd \; \left[\;\left[\;x_i,{x}_{j}\;\right]_{\star_{\mathcal{F}}}-iw(t)(\delta_{1i}\delta_{2j} - \delta_{1j}\delta_{2i})\;\right] = 0\;, \label{ffrow4}\\
M_{13} &\rhd& \left[\;\left[\;x_i,{x}_{j}\;\right]_{\star_{\mathcal{F}}}-iw(t)(\delta_{1i}\delta_{2j} - \delta_{1j}\delta_{2i}) \;\right] = w(t)(\delta_{2i}\delta_{3j}-\delta_{2j}\delta_{3i})\;, \label{ffrow5}\\
M_{23} &\rhd& \left[\;\left[\;x_i,{x}_{j}\;\right]_{\star_{\mathcal{F}}}-iw(t)(\delta_{1i}\delta_{2j} - \delta_{1j}\delta_{2i}) \;\right] = -w(t)(\delta_{1i}\delta_{3j}-\delta_{1j}\delta_{3i})\;, \label{ffrow6}\\
H &\rhd& \left[\;\left[\;x_i,{x}_{j}\;\right]_{\star_{\mathcal{F}}}-iw(t)(\delta_{1i}\delta_{2j} - \delta_{1j}\delta_{2i}) \;\right] = g(t)(\delta_{1i}\delta_{2j} - \delta_{1j}\delta_{2i})\;,\\
&&~~~~~~~~~~~~~~~~~~~~~~~~~~~~~~~~~~~~~~~~~H  \;\rhd\; \left[\;t,{x}_{i}\;\right]_{\star_{\mathcal{F}}}  = 0\;,\label{ffrow5}
\end{eqnarray}
where $(g_{\kappa_i}(t) = \lim_{\tau \to \infty} h_{\kappa_i}(t))$
\begin{eqnarray}
g({t})&=&g_{\kappa_1}({t}) =
0\;, \label{oopw2}\\
g({t})&=&g_{\kappa_2}({t}) =
\kappa_2 \;,
\label{oopw3}\\
g({t})&=&g_{\kappa_3}({t}) =
2\kappa_3t \;, \label{oopw4}\\
g({t})&=&g_{\kappa_4}({t}) =
 4\kappa_4
 t^3 \;, \label{oopw5}\\
g({t})&=&g_{\kappa_5}({t}) =
\kappa_5t\;, \label{oopw6}\\
g({t})&=&g_{\kappa_6}({t}) =
 \frac{3}{2}\kappa_6 t^2\;.\label{oopw7}
\end{eqnarray}
The above result means that the commutations relations (\ref{wwspaces}) remain invariant with respect the action of
$P_i$, $K_i$, $F_i$, $M_{12}$ and $H$ generators in the case of deformation (\ref{ggnw2}) as well as
$P_i$, $K_i$, $F_i$ and $M_{12}$ for space-times (\ref{ggnw3})-(\ref{ggnw7}).

Finally, let us observe that the above considerations can be applied to the case of classical Galilei quantum
group $\;{\mathcal U}_0({G})$ by neglecting operators $F_i$.

\section{Final remarks}

In this article we provide the subgroups of classical acceleration-enlarged Newton-Hooke
$\;{\mathcal U}_0(\widehat{ NH}_{\pm})$ as well as classical acceleration-enlarged Galilei $\;{\mathcal U}_0(\widehat{G})$ Hopf
structures, which play the role of "isometry" groups for
twist-deformed space-times (\ref{spaces}) and (\ref{wwspaces}). In such a way, by analogy to the investigations
performed in \cite{link1}, \cite{link2}, we get the link between
twisted quantum spaces and the proper undeformed Hopf subalgebras.
Consequently, the obtained results admit to analyze the
twist-deformed dynamical models \cite{model0}-\cite{model2} in terms
of the corresponding classical quantum subgroups of the whole  nonrelativistic symmetries. The works in this direction already
started and are  in progress.

\section*{Acknowledgments}
The author would like to thank J. Lukierski
for valuable discussions.

\end{document}